\documentclass[smallextended,final]{svjour3} 

\usepackage[pdftex]{graphicx}       
\usepackage{amsmath}
\usepackage{amssymb}
\usepackage[numbers,sort&compress]{natbib}
\newcommand{\avm}{\langle m \rangle / n}
\newcommand{\avs}{\langle s \rangle / n}
\newcommand{\avz}{\langle \zeta \rangle}
\newcommand{\avR}{\langle R_e^2 \rangle}
\newcommand{\eref}[1]{Eq.~\eqref{#1}}
\newcommand{\fref}[1]{Fig.~\ref{#1}}

\usepackage{hyperref} 
\hypersetup{colorlinks,citecolor=blue,filecolor=blue,linkcolor=blue,urlcolor=blue}

\begin{document}

\title{Effect of lattice inhomogeneity on collapsed phases of semi-stiff ISAW polymers}

\author{C. J. Bradly \and A. L. Owczarek} 
\institute{School of Mathematics and Statistics, University of Melbourne, Victoria 3010, Australia \\ \email{chris.bradly@unimelb.edu.au}}
\date{\today}

\maketitle

\begin{abstract}
We investigate semi-stiff interacting self-avoiding walks on the square lattice with random impurities.
The walks are simulated using the flatPERM algorithm and the inhomogeneity is realised as a random fraction of the lattice that is unavailable to the walks.
We calculate several thermodynamic and metric quantities to map out the phase diagram and look at how the amount of disorder affects the properties of each phase.
On a homogeneous lattice this model has an extended phase and two distinct collapsed phases, globular and crystalline, which differ in the anisotropy of the walks.
By adding impurities to the lattice we notice a degree of swelling of the walks for all phases that is commensurate with the fraction of the lattice that is removed.
Importantly,  the crystal phase disappears with the addition of impurities for sufficiently long walks. 
For finite length walks we demonstrate that competition between the size of the average spaces free of impurities and the size of the collapsed polymer describes the crossover between the homogeneous lattice and the impurity dominated situation.
\keywords{polymer collapse, inhomogeneous lattice, self-avoiding walks}
\end{abstract}

\section{Introduction}
\label{sec:Intro}

The canonical picture of polymer collapse is that long chain polymers in a poor solvent undergo a phase transition from an extended phase to a collapsed phase at a critical temperature, known as the $\theta$-point \cite{Flory1953,Duplantier1987}.
In the interacting self-avoiding walk (ISAW) model this is induced by an interaction between non-consecutive monomers causing the polymer to energetically favour a random globule configuration at low temperatures.
There are, however, a number of other lattice models of polymer collapse with different properties which may suggest the $\theta$-point transition is not universal \cite{Blote1989,Shapir1984}.
Often these models involve a change to the self-avoiding condition of the lattice walk such as the vertex-interacting SAW \cite{Foster2003}, or the self avoiding trail \cite{Owczarek1995,Owczarek2007a} but can also be due to the addition of another interaction \cite{Doukas2010,Bedini2016}.
In the latter case a new collapsed phase can appear where the configurations of the walk have crystalline structure.
A similar example of this kind of model is the semi-stiff ISAW \cite{Bastolla1997,Krawczyk2009} which allows for varying the energy cost of a bend in a walk.
When bends are inhibited the typical configurations resemble $\beta$-sheet configurations and the transition between the extended phase and the crystal phase is first order, compared to the continuous $\theta$-point transition to the random globule phase in the flexible regime.

If the collapsed phase is characterised by the formation of dense configurations then it is sensible to ask how robust this phase is to some degree of inhomogeneity in the underlying medium.
The effect of random disorder on otherwise well-understood statistical mechanical problems has been a topic of study for some time, going back to the Ising model \cite{Watson1969}.
The study of the SAW model on randomly diluted lattices goes back almost as far and has been closely related to the problem of percolation.
By modeling disorder in the underlying medium as percolation early studies showed that the fundamental scaling laws for SAWs, that of the size and number of walks, have the same universal exponents $\nu$ and $\gamma$ on inhomogeneous lattices as for homogeneous lattices, provided the disorder is above the percolation limit $p_c$ \cite{Kremer1981}.
Change in this scaling behaviour only occurs at the percolation limit $p_c$ \cite{Blavatska2008}.
These results have been confirmed with numerical work \cite{Lee1988,Rintoul1994} and exact enumeration \cite{Lam1990,Ordemann2000,Nakanishi1991}. 
Renormalisation group theory also finds that \cite{Kim1983,Blavatska2005} however there is evidence of additional fixed points due to the disordered medium \cite{Meir1989,Grassberger1993}.
The addition of disorder also introduces new considerations such as how the type of averaging over disorder affects SAWs \cite{Nakanishi1992,Birkner2010} and when scaling laws are well-defined \cite{Janssen2007}.
Other approaches to linear polymers in disordered media include fractal and irregular lattices \cite{Chakrabarti2005,Dhar2005,Rammal1984,Roy1990}.

The link back to polymer collapse involves the formation of sub-clusters in a SAW on a smaller scale which can be related to cluster formation in percolation models \cite{Brak1998} and different scaling if one considers SAWs on a single percolation cluster or across multiple percolation clusters \cite{Sahimi1984}.
It is also known that the (effective) connective constant of SAWs on a regular lattice decreases linearly with bond dilution $1-p$ but the $\theta$-point is only weakly affected, even down to critical $p_c$ \cite{Barat1991}.
In this paper we use the percolation formulation as the representation of lattice inhomogeneity, where each site has a probability $p$ to be open.
From the point of view of a lattice polymer, each site has a probability $1-p$ to be an invalid step in the walk, thus serving as an impurity.
We are more interested in this work on the effect of the inhomogeneity on the collapsed phase of the semi-stiff ISAW model so we work well above the percolation limit $p_c \approx 0.593$ \cite{Newman2000} on the square lattice where scaling laws are unchanged.
Away from the fixed points at $1-p = 0, 1-p_c$ the disruption to the symmetry of the lattice should have a significant effect on the highly ordered crystal phase.
We characterise the phases present in this system and see how they are affected by an increase in lattice inhomogeneity.
In particular, we investigate whether the crystal phase is more sensitive to disruption by impurities compared to the globule phase.

\section{Model and simulation}
\label{sec:Model}

\begin{figure}[t!]
	\centering
	\includegraphics[width=0.5\columnwidth]{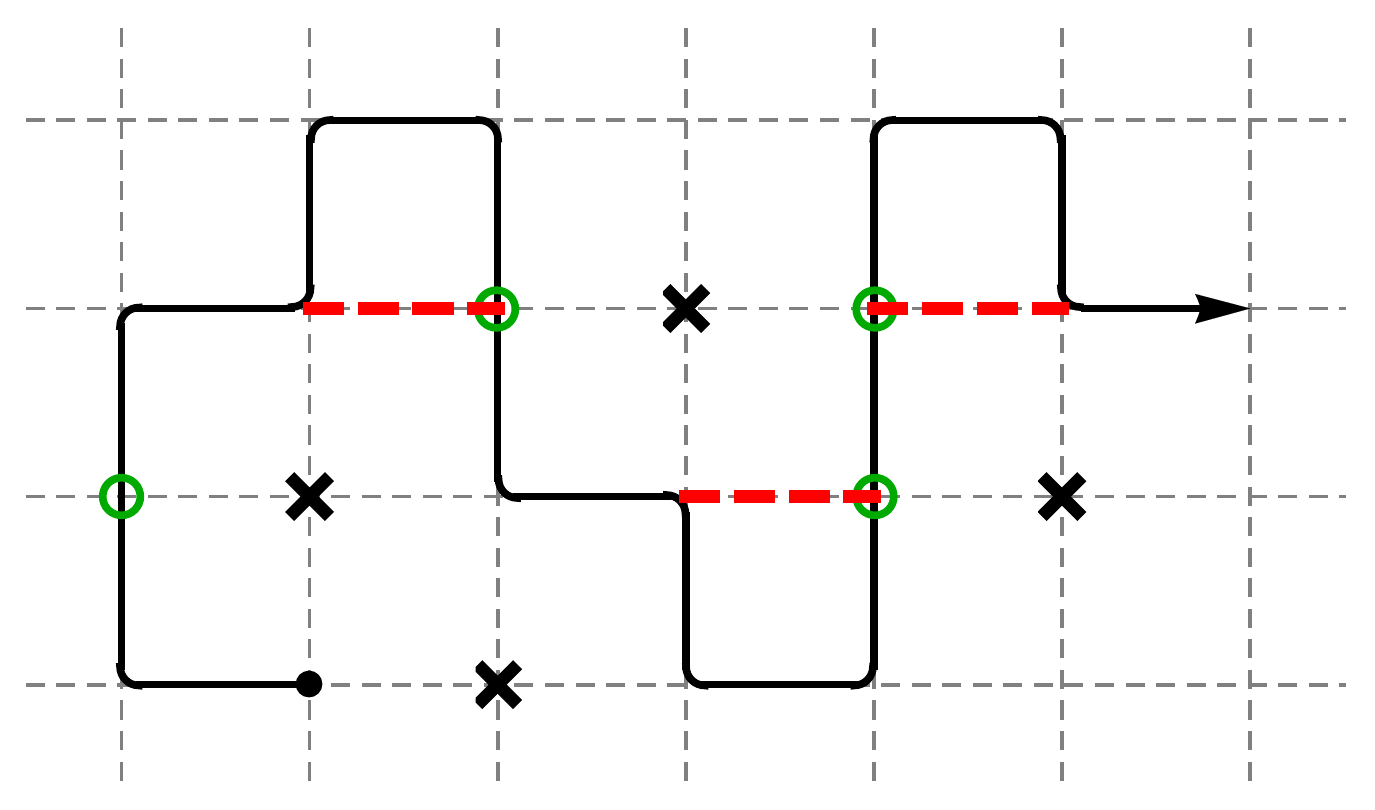}
	\caption{A sample of a SAW on the square lattice with $s=4$ straight segments (green circles) and $m=3$ near-neighbour interactions (red dashed lines). 
	Impurities in the lattice are marked with black crosses.}
	\label{fig:InhomoSSISAW}
\end{figure}

We consider single polymers in dilute solution modeled as self-avoiding walks (SAWs) on the square lattice.
The canonical partition function for walks of length $n$ with $m$ bulk interactions and $s$ straight segments is
\begin{equation}
	Z_n = \sum_{m,s} c_{n}(m,s) \, \omega^m \eta^s,
	\label{eq:CombinedPartition}
\end{equation}
where $\omega$ and $\eta$ are the Boltzmann weights of a non-consecutive neighbouring pair and a straight segment, respectively, and $c_{n}(m,s)$ is the density of states.
To represent the presence of impurities, each lattice site has a probability $p$ that it is a valid site for a step in the walk.
Thus the inhomogeneous lattice is represented by making a fraction $1-p$ of lattice sites unavailable.
An example walk is shown in \fref{fig:InhomoSSISAW}.
Details of how the lattice configuration is chosen are given below when discussing the flatPERM algorithm.

To characterise the phases of the system, particularly the collapsed phases, we calculate the internal energy-type order parameters, namely the average fraction of straight segments $\avs$ and the average fraction of neighbouring steps $\avm$.
An additional signature of the crystal phase is the anisotropy, defined as
\begin{equation}
	\zeta = 1 - \frac{\min(n_x,n_y)}{\max(n_x,n_y)},
\label{eq:Anisotropy}
\end{equation}
where $n_x$ and $n_y$ are the number of steps of the walk in each orthogonal direction of the lattice.
In the case of a homogeneous lattice the collapsed phases are indicated by a non-zero value of $\avz$ with $\avz =1$ in the crystal phase.
Within the collapsed regime we expect $\avs$ to change value across the globule-crystal transition while $\avm$ remains smooth and varies slowly.
The effect of lattice inhomogeneity on the system will be seen in how these quantities vary with $1-p$.

The effect of disorder on the nature of the phase transitions determined from scaling laws.
In particular, in the thermodynamic limit of long chains the specific heat near the transition temperature $T_c$ scales according to
\begin{equation}
	c(T) \sim \left| T - T_c \right|^{-\alpha},
\label{eq:SpecificHeat}
\end{equation}
controlled by the critical exponent $\alpha$.
If $\alpha < 1$ the transition is continuous; if $\alpha = 1$ then \eref{eq:SpecificHeat} diverges, indicating a first-order transition.
For a finite system we calculate the reduced specific heat from moments of the partition function, for example
\begin{equation}
	c_n^{(m)}(\omega) = \frac{\text{var}(m)}{n} = \frac{\langle m^2 \rangle - \langle m \rangle^2}{n},
\label{eq:VarM}
\end{equation}
and similar for $c_n^{(s)}(\eta) = \text{var}(s)/n$.
The peak of $c_n^{(m)}$ is at the transition point $\omega_c$ and the value of the peak scales as $n^{\alpha \phi}$, where $\phi$ is the crossover exponent.
By considering \eref{eq:VarM} and using the standard relation $2-\alpha = 1/\phi$ we may determine the nature of the transitions.
The rule of thumb is that if the peak values of $c_n^{(m)}/n$ coincide for a range of $n$ then $\alpha \phi = 1$, thus $\alpha = 1$ and the transition is first-order; else, continuous.
Since we have two microcanonical parameters we generalise and calculate the Hessian covariance matrix
\begin{equation} 
	H_n =
	\begin{pmatrix}
	 \frac{\partial^2 f_n}{\partial \omega^2} 	& \frac{\partial^2 f_n}{\partial \omega \partial \eta}	\\
	 \frac{\partial^2 f_n}{\partial \eta \partial \omega} 	& \frac{\partial^2 f_n}{\partial \eta^2}
	\end{pmatrix}
	,
	\label{eq:Hessian}
\end{equation}
where $f_n = -\tfrac{1}{n}\log Z_n$ is the reduced free energy.

	
Walks are simulated using the flatPERM algorithm \cite{Prellberg2004}, an extension of the pruned and enriched Rosenbluth method (PERM) \cite{Grassberger1997}. 
The algorithm works by growing a walk on a given lattice up to some maximum length $N_\text{max}$. 
At each step the number of bulk interactions $m$ and straight segments $s$ are calculated and the cumulative Rosenbluth \& Rosenbluth weight \cite{Rosenbluth1955} is compared with the current estimate of the weights of all samples $W_{n,m,s}$. 
If the current state has relatively low weight the walk is `pruned' back to an earlier state. 
On the other hand, if the current state has relatively high weight, then microcanonical quantities $m$ and $s$ are measured and $W_{n,m,s}$ is updated. 
The state is then `enriched' by branching the simulation into several possible further paths (which are explored when the current path is eventually pruned back). 
When all branches are pruned a new iteration is started from the origin.
FlatPERM improves upon PERM by altering the prune or enrich choice such that the sample histogram is flat in the microcanonical parameters $n$, $m$ and $s$. 
Further improvements are made to account for the correlation between branches that are grown from the same enrichment point, which provides an estimate of the number of effectively independent samples. 
The main output of the simulation are the weights $W_{n,m,s}$, which are an approximation to the athermal density of states $c_{n}(m,s)$ in \eref{eq:CombinedPartition}, for all $n\le N_\text{max}$. 
In practice, thermodynamic quantities are determined by specifying $\omega$ and $\eta$ and using the weighted sum
\begin{equation}
    \langle Q \rangle(\omega,\eta) = \frac{\sum_{m,s} Q_{m,s} \omega^m \eta^s W_{n,m,s}}{\sum_{m,s} \omega^m \eta^s W_{n,m,s}}.
    \label{eq:FPQuantity}
\end{equation}
For each case or simulation mentioned in this work the results are comprised of a composite of ten independent simulations in order to obtain some measure of statistical error.

The inhomogeneous lattice is implemented by choosing a set of lattice sites to be inaccessible to the walk.
The number of impurities is drawn from the appropriate binomial distribution with $1-p$ being the probability of any particular site being a valid site for the walk.
These impurities are distributed uniformly over the area of the lattice that would be accessible to a walk of length $n$ in the case of a homogeneous lattice.
The set of inaccessible sites is reseeded at the beginning of each flatPERM iteration (growing the walk from the origin).
The initial weight of each iteration is set to be the probability of the configuration of lattice impurities.
In this way the output weights $W_{n,m,s}$ contain the sum over disorder such that any $\langle Q \rangle$ in \eref{eq:FPQuantity} also represents a quenched-type average over disorder \cite{Nakanishi1992}.
It is also possible to achieve annealed-type averages by ignoring the probability of the lattice configuration at each iteration but we found no appreciable difference between the two methods for this problem.

\begin{figure}[t!]
\centering
\includegraphics[width=\textwidth]{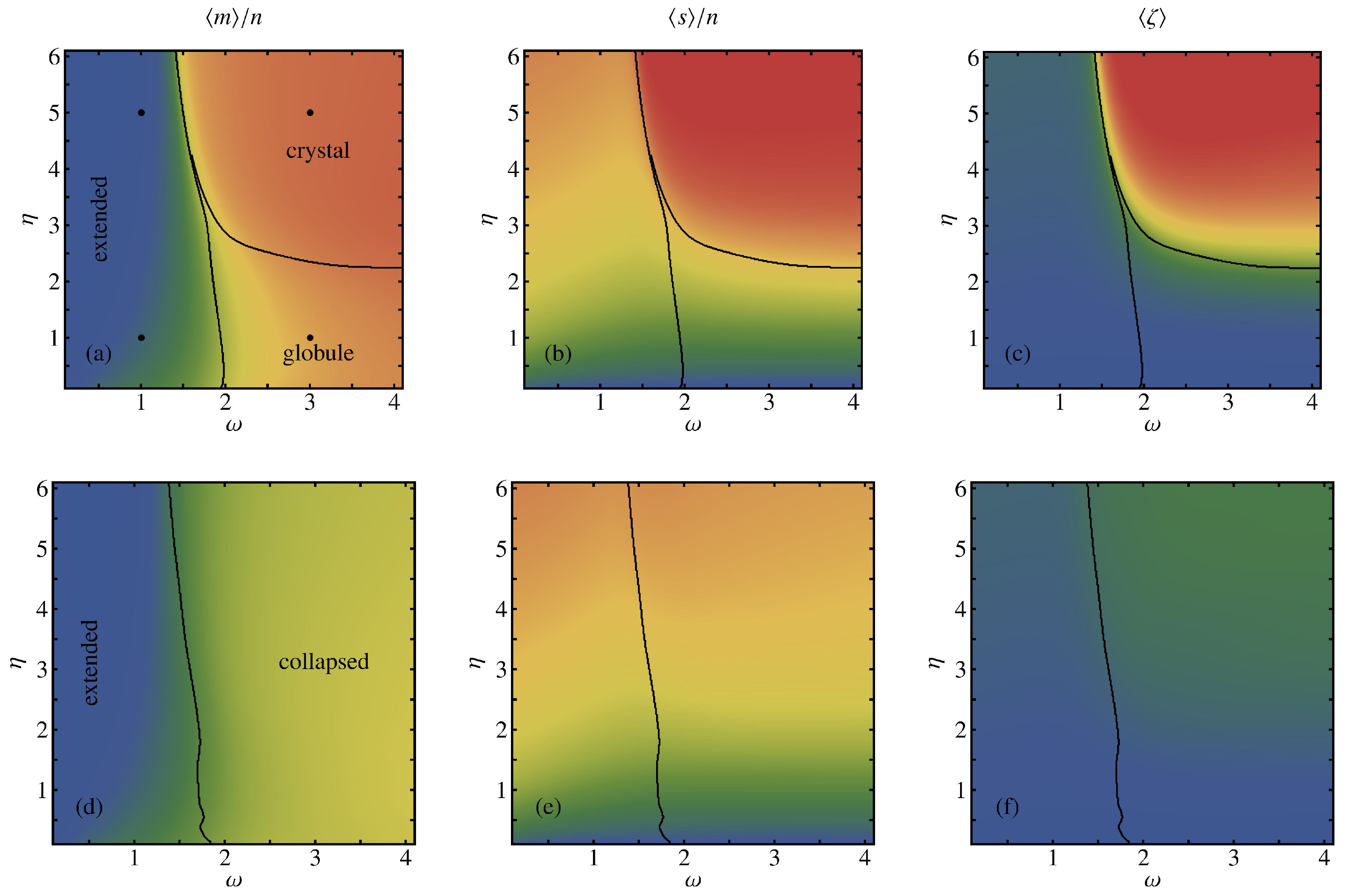}
\caption{The order parameters (from left to right) average fraction of neighbouring steps $\avm$, average fraction of straight segments $\avs$ and average anisotropy $\avz$. 
Plots are for length $n = 256$ and $1-p = 0$ (a-c, top) and $1-p = 0.25$ (d-f, bottom).
In all cases the scale is from 0 (blue) to 1 (red).
Black lines are schematic phase boundaries as determined by points of peak variance; see \fref{fig:HessianPhase}.}%
\label{fig:OrderParameters}%
\end{figure}

\begin{figure}[t!]
\centering
\includegraphics[width=0.8\textwidth]{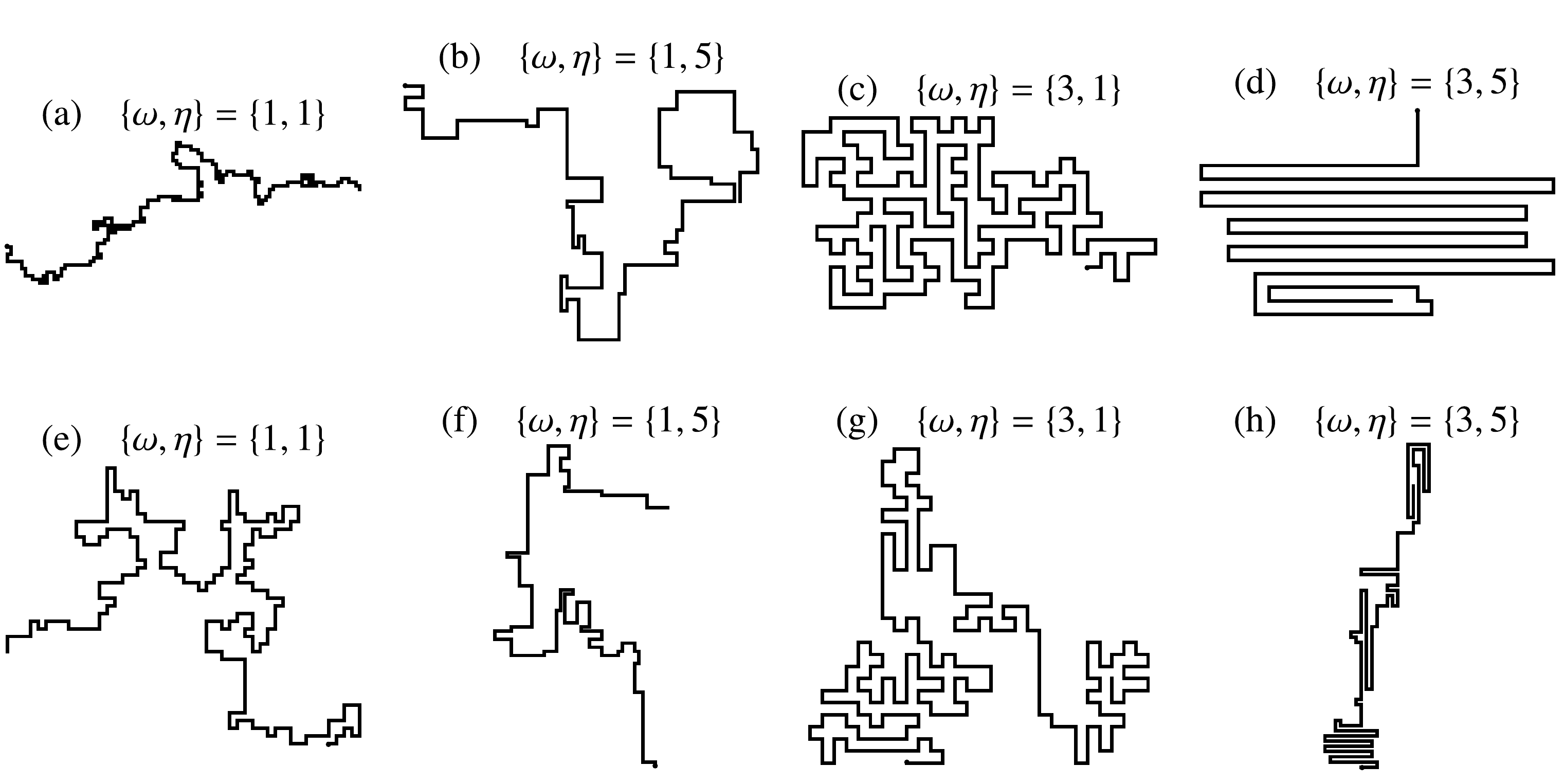}
\caption{Typical configurations at points in the phase diagram indicated on \fref{fig:OrderParameters}. Top row (a-d) $1-p = 0$ and Top bottom row (e-h) $1-p = 0.25$.}%
\label{fig:Configurations}%
\end{figure}

\section{Phase diagram}
\label{sec:Phase}

First we characterise the phases by looking at the order parameters for the system and the expected configurations, with and without lattice impurities.
In \fref{fig:OrderParameters} we plot, from left to right, the average fraction of straight segments $\avs$, the average fraction of neighbouring steps $\avm$ and the average anisotropy $\avz$.
Further visualisation of the phases is given in \fref{fig:Configurations} which shows typical configurations at points in the $(\omega,\eta)$ phase diagram that are indicative of each phase.
These points are marked with black dots on \fref{fig:OrderParameters}(a).
In both figures the top and bottom row are for $1-p = 0,0.25$, respectively.
In the homogenous lattice case, we see the three distinct phases of the semi-stiff ISAW system \cite{Krawczyk2009}.
For small $\omega$ at all $\eta$ the chains are extended, with near-zero fraction of neighbouring monomers arranged isotropically and the fraction of straight segments increasing weakly with $\eta$.
Configurations in this phase look like \fref{fig:Configurations}(a,b).
For small $\eta$ and $\omega > \omega_\text{c}$ the chains are collapsed in a random (and thus isotropic) globule, with a small fraction of straight segments, like in \fref{fig:OrderParameters}(c).
At $\eta = 1$ the chains are completely flexible and there a collapse transition at $\omega = \omega_\theta$ so that for large $\omega > \omega_c$ the polymer is in an amorphous dense state where the average size scales as $\sqrt{n}$.
The third phase is the crystal phase which is also collapsed since where the average size scales as $\sqrt{n}$ which is characterised by folded configurations of zero entropy. This phase occurs when $\eta$ is larger than some critical value $\eta_c$ such that the fraction of straight segments is high, and thus the chains have anisotropic conformations arranged as parallel lines, like in \fref{fig:OrderParameters}(d), analogous to $\beta$-sheets in three dimensions.

In the case of the inhomogeneous lattice, where a considerable fraction of the lattice is unavailable to the walks, there are several differences.
The extended phase ($\omega < \omega_c$) is still clear in this case, marked by small but non-zero $\avm$, $\avz \approx 0$ and $\avs$ increasing weakly with $\eta$, as in the homogeneous case.
The configurations in \fref{fig:Configurations}(e,f) are like those in \fref{fig:Configurations}(a,b).
A collapsed phase still exists for $\omega > \omega_c$, as evidenced in \fref{fig:OrderParameters}(d) where the value of $\avm$ is non-zero, albeit smaller than the homogeneous lattice case.the globule and crystal phases.
This is noticeable in \fref{fig:OrderParameters}(e) where the value of $\avs$ is reduced for the inhomogeneous lattice in the high stiffness and nearest neighbour interaction region ($\omega > \omega_c$ and $\eta > \eta_c$).
Moreover, $\avs$ increases only slowly with increasing $\eta$ and is largely independent of $\omega$ in this region.
This suggests a smooth crossover as $\eta$ increases for $\omega > \omega_c$, rather than a phase transition as is the case for the homogeneous lattice.
Similarly, $\avz$ is greatly reduced in this region, however it remains non-zero, as shown in \fref{fig:OrderParameters}(f).
The corresponding configurations in \fref{fig:Configurations}(g,h) - for the globule, crystal phase, respectively - show a change in the global structure (compared to \fref{fig:Configurations}(c,d)), with the appearance of several connected smaller clusters.
In particular, the straightness of the connecting strands between clusters accounts for the small but non-zero value of $\avz$.

\begin{figure}[t!]
\centering
\includegraphics[width=0.5\columnwidth]{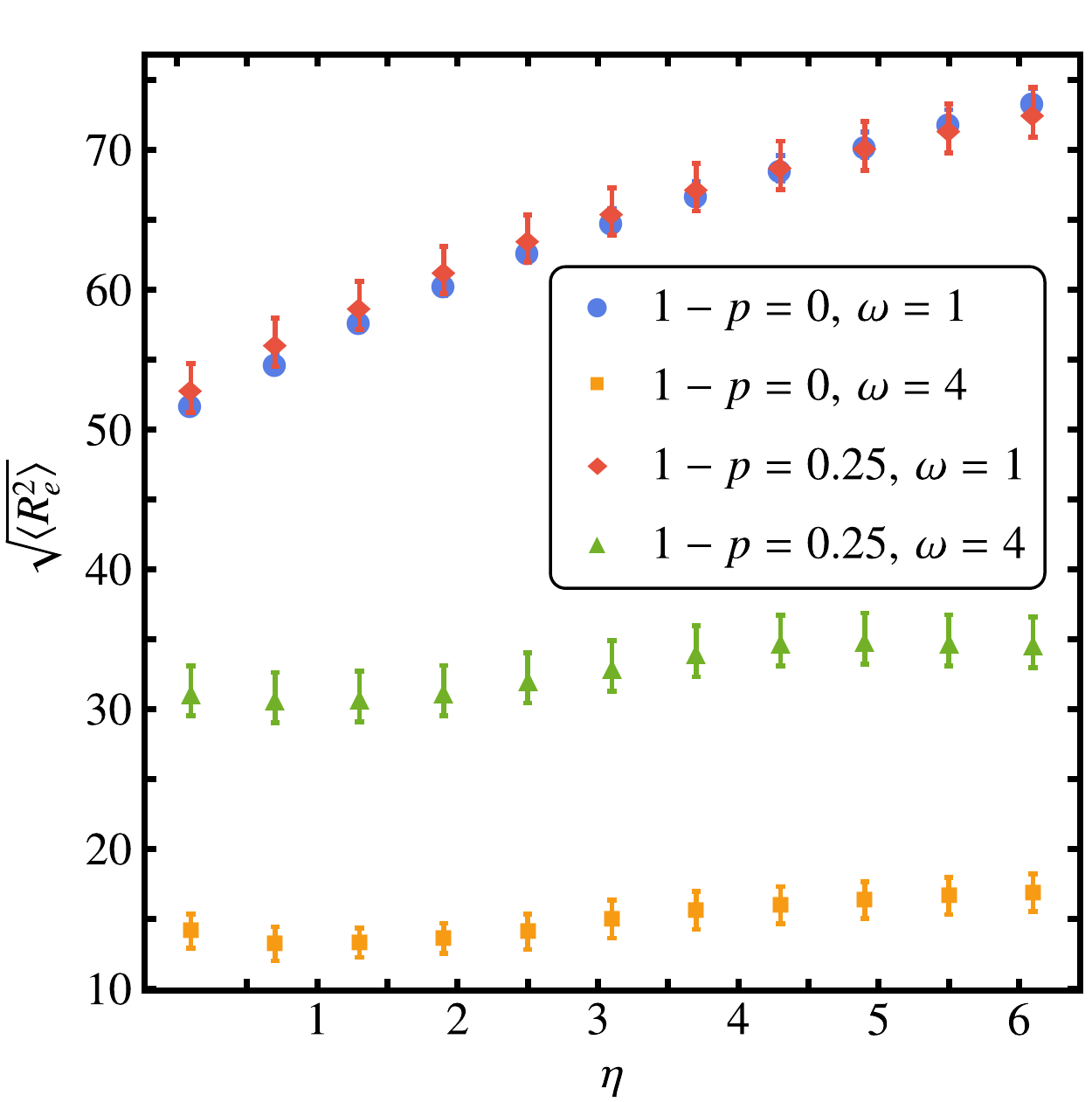}
\caption{The (square-root of) the mean-squared end-to-end distance $\avR$ with and without lattice inhomogeneity in the extended phase ($\omega=1$) and the collapsed phases ($\omega = 4$).
Data is from the $n=256$ simulations.}%
\label{fig:R2Slices}%
\end{figure}

Another quantity of interest for polymers is the mean-squared end-to-end distance $\avR$, where for a single chain $R_e^2 = \mathbf{r}_n.\mathbf{r}_n$
assuming the chain starts at the origin $\mathbf{r}_0 = (0,0)$.
In \fref{fig:R2Slices} we plot $\avR$ for two regions of the phase diagram, $\omega = 1$ and $\omega = 4$ corresponding to the extended phase and the collapsed phases, respectively, as well as with and without lattice inhomogeneity, $1-p = 0$ and $1-p = 0.25$, respectively.
The top two series (blue and red) are for the extended phase and show only a small difference between the homogeneous and inhomogeneous lattice.
This small difference suggests a slight swelling of the chain for small $\eta$ when inhomogeneity reduces the number of available sites for the chain.
Interestingly, this difference reverses at larger $\eta$ where samples of very long straight chains are inhibited by the inhomogeneous lattice.
There is a larger distinction between the homogeneous and inhomogeneous lattice cases in the collapsed phases (green and orange).
On the inhomogeneous lattice the chains are significantly larger than on on homogeneous lattice though presumably still obeying $\sqrt{\avR} \sim \sqrt{n}$; our data is not sufficient to verify this scaling.
However, apart from a small effect due to the homogeneous lattice globule-crystal transition near $\eta \approx 2$, the size of the chains are not significantly affected by stiffness in the collapsed phases for both the homogeneous and inhomogeneous lattices.

\begin{figure}[t!]
\centering
\includegraphics[width=0.8\columnwidth]{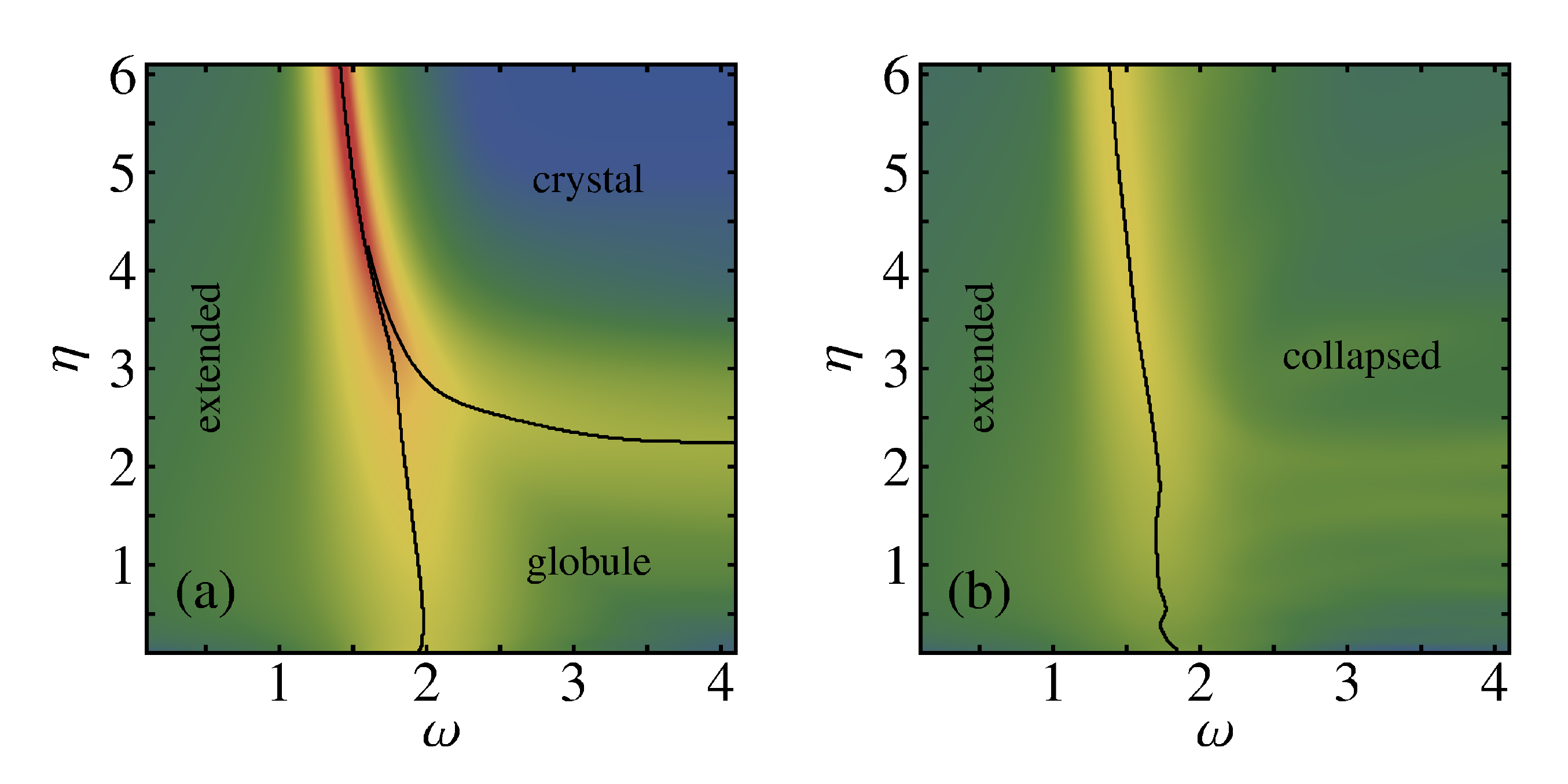}
\caption{The largest eigenvalue of the Hessian matrix, representing covariance of $m$ and $s$.
Points of peak covariance are marked with black lines and serve as schematic phase boundaries.
Data is from a simulation up to length $n=256$ for (a) $1-p = 0$ and (b) $1-p = 0.25$.}%
\label{fig:HessianPhase}%
\end{figure}

\begin{figure*}[t!]
\centering
\includegraphics[width=0.8\textwidth]{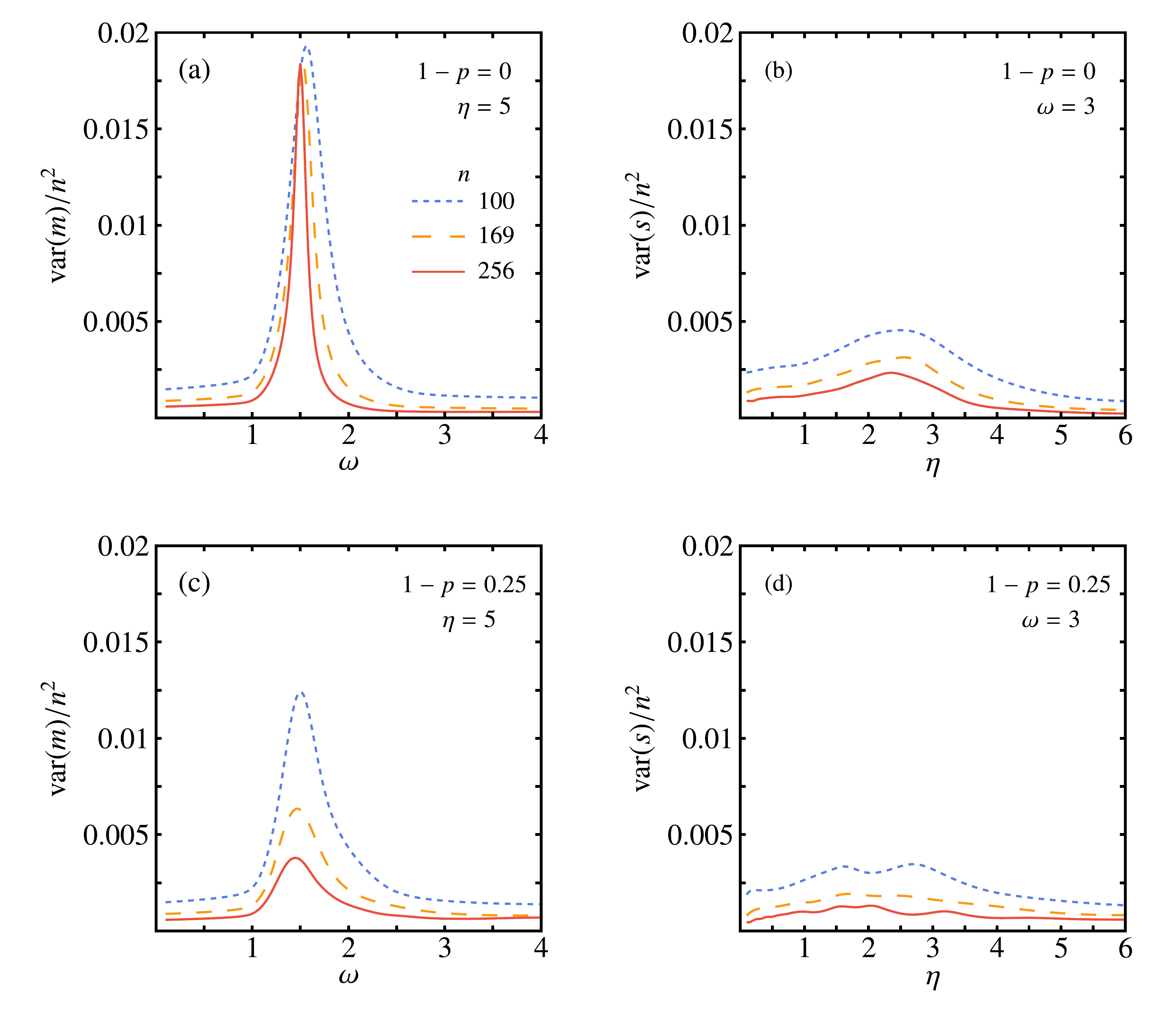}
\caption{Variance of $m$ and $s$ scaled by $n^2$ across phase transition boundaries for a few values of $n$.
(a) and (c) are for a slice at fixed $\eta = 5$ across the extended-crystal transition, with $1-p = 0,0.25$, respectively.
(b) and (d) are for a slice at fixed $\omega= 5$ across the globule-crystal transition, with $1-p = 0,0.25$, respectively.
}%
\label{fig:VariancePeakScaling}%
\end{figure*}

The effect of lattice inhomogeneity on the phase boundaries can be seen more clearly in \fref{fig:HessianPhase} which plots the largest eigenvalue of the covariance matrix $H_n$ of \eref{eq:Hessian}.
For $p = 0$ (left) the phase boundaries are clear, especially between the extended and crystal phases.
The value of $\omega_c$ that marks this boundary decreases slightly as $\eta$ increases, but corresponds to the $\theta$-point at
$\omega_\theta \approx 1.94 $ for SAWs on the square lattice \cite{Meirovitch1989}.
In contrast, lattice impurities ($1-p = 0.25$, right) soften the boundaries.
While the variance that marks the boundary between the extended and collapsed phases is smaller it is still clear. 
To determine the nature of these transitions we look at slices of \fref{fig:HessianPhase} at fixed values of $\omega$ or $\eta$.
Away from the multicritical point, off-diagonal terms of the Hessian matrix are negligible and so the largest eigenvalue is equivalent to the largest specific heat.
Thus, in \fref{fig:VariancePeakScaling} we plot the specific heats scaled by $n$ to determine the scaling exponent $\alpha$.
The (scaled) variance of $m$ across the extended-crystal transition is shown in \fref{fig:VariancePeakScaling}(a) and (c) at fixed $\eta = 5$ for $1-p = 0$ and $1-p = 0.25$, respectively.
On the homogeneous lattice (a), the peaks of $\text{var(m)}/n^2$ coincide, indicating $\alpha \approx 1$, consistent with a first-order transition.
In the presence of disorder (c), the scaling indicates $\alpha < 1$, and thus the transition is now continuous.
This is a clear example of quenched disorder causing the rounding of a first-order transition in 2D \cite{Hui1989,Aizenman1989}.
For the globule-crystal transition at fixed $\omega = 3$ we show the variance of $s$ scaled by $n$, again for $1-p = 0$ and $1-p = 0.25$, in \fref{fig:VariancePeakScaling}(b) and (d), respectively.
For $1-p = 0$ (b) the variance of $s$ indicating a globule-crystal transition is broader and smaller than the extended-crystal transition, but nevertheless it is clear enough to indicate a continuous transition, with $\alpha < 1$.
However, in the presence of disorder (d), there is no clear signal of a transition at all.
Along with the effect on the order parameters and $\avR$, we can confidently argue that that the distinction between the globule and crystal phase regions no longer exists for SAWs on an inhomogeneous lattice. 
We conclude that the crystal phase only truly exists for $p=1$.

\begin{figure*}[t!]
\centering
\includegraphics[width=\textwidth]{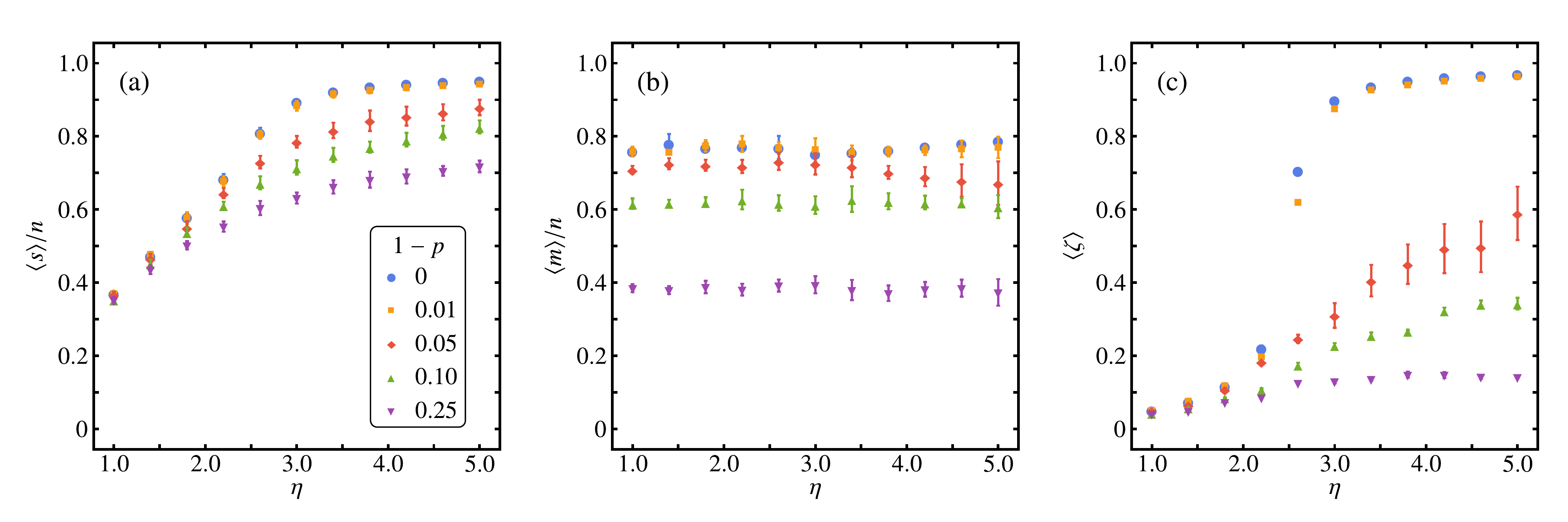}
\caption{Order parameters (a) average fraction of straight segments $\avs$, (b) average fraction of neighbouring steps $\avm$ and (c) average anisotropy $\avz$. 
Plots are for length $n = 1024$, $\omega = 3$ and a range of impurity densities $1-p$. }%
\label{fig:OrderParametersSlice}%
\end{figure*}

So far we have compared the homogeneous lattice case to a case with a significant amount of inhomogeneity. 
While the value $1-p = 0.25$ is not close to the percolation limit $1-p_c$ it still makes a significant fraction of the lattice unavailable to the walk. 
We now turn our attention to the scaling region around $1-p \to 0$ and $n$ large. 
For a better look at this scaling region  we also ran a set of additional simulations with a fixed interaction weight $\omega = 3$ corresponding to a slice in the phase diagram that avoids the extended phase and looks for and change in the collapsed phase as $\eta$ is increased.
This value of $\omega$ is large enough to avoid the collapse transition without being overwhelmingly biased towards collapsed configurations.
This is achieved by altering the weights of the samples by a factor $\omega^m$ so that the sum over $m$ in \eref{eq:FPQuantity} is performed within the simulation.
This method produces histograms over $n$ and $s$ only and can sample longer walks up to $n = 1024$.
Figure \ref{fig:OrderParametersSlice} shows (a) average fraction of straight segments $\avs$, (b) average fraction of neighbouring steps $\avm$ and (c) average anisotropy $\avz$.
As expected, increasing $\eta$ is marked by a change in the behaviour of $\avs$ and $\avz$, but not $\avm$.
This is broadly true for all values of $1-p$ but other effects due to increasing $1-p$ are also evident.
For $1-p$ near or equal to zero $\avs$ has an approximately linear $\eta$ dependence at small values of $\eta$ and $\eta \approx 2.5$ $\avs$ tends to a constant value.
This is the continuous globule-crystal transition seen earlier and we see that it softens considerably for larger $1-p$ where there is no crystal phase.
At small $1-p$ the limiting value at large $\eta$ is close to unity and it decreases slightly as $1-p$ increases.
This indicates that although the available fraction of the lattice may diminish it is still possible for configurations to have large straight segments and $\avs$ does not necessarily say anything about the mean size of the chains; compare the configurations in \fref{fig:Configurations}(d,h), which have similar values of $s/n$.
In \fref{fig:OrderParametersSlice}(b) we also see that $\avm$ decreases smoothly with increasing $1-p$ at all values of $\eta$.
However, the effect is stronger compared to $\avs$ since each impurity prevents multiple near-neighbour bonds, presumably as a function of the coordination number of the lattice.
This corresponds partly to a swelling in the size of the chain but also to the formation of smaller clusters, as seen in \fref{fig:Configurations}(g).
Although we know that the crystal-globule transition has disappeared, it is not clear from the plots of $\avs$ and $\avm$ whether there is a significant change in the properties of the collapsed phase in the large $\eta$ region as the disorder increases.

\begin{figure}[t]
\centering
\includegraphics[width=0.5\columnwidth]{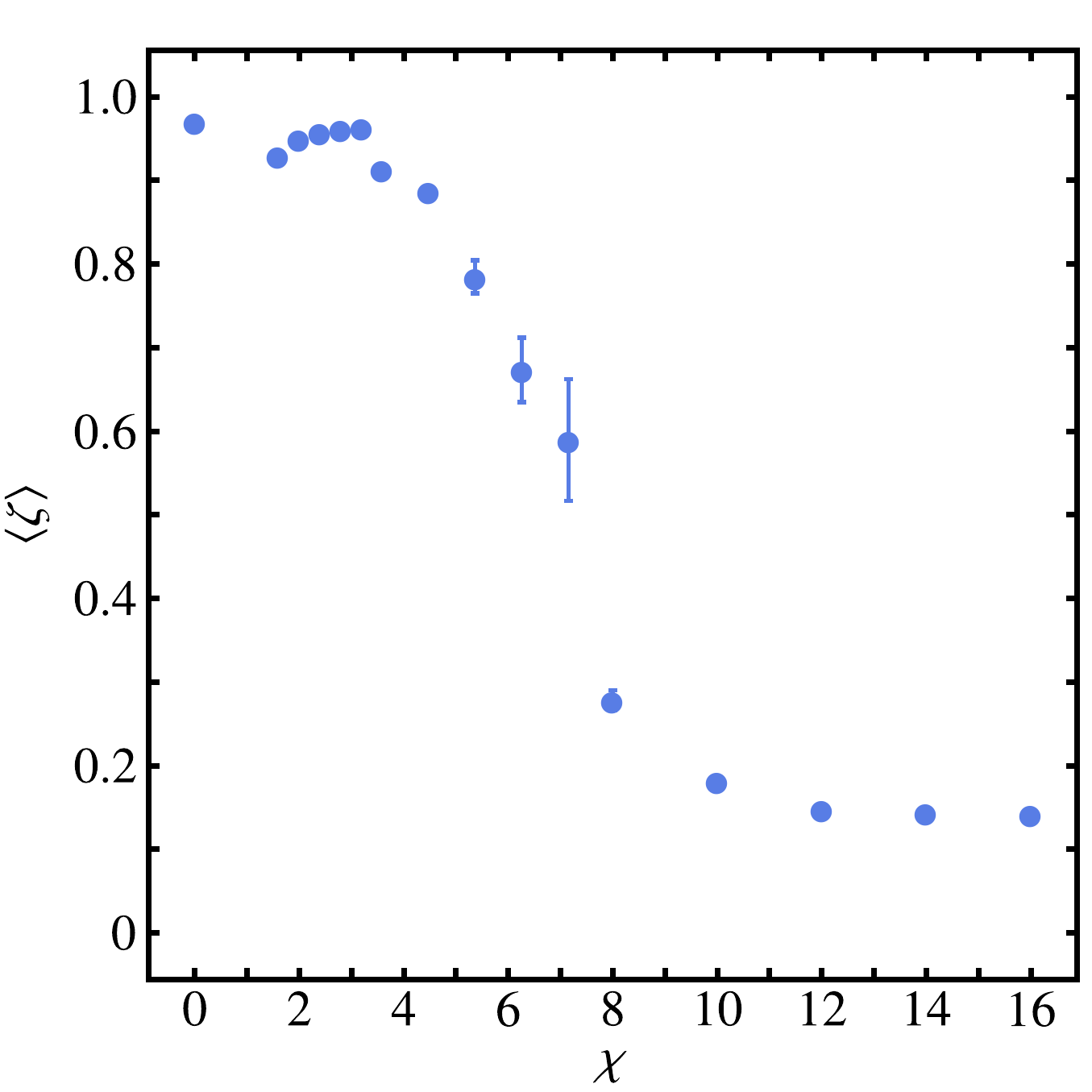}
\caption{The low-temperature ($\eta = 5$), large $n$ limiting values of $\avz$ as a function of $\chi = \sqrt{n \, (1-p)}$. 
For each $1-p$ we use a range $256 < n < 1024$ to achieve different values of $\chi$.}
\label{fig:LimitingOrderParameters}%
\end{figure}

The more interesting behaviour is in the anisotropy $\avz$ in \fref{fig:OrderParametersSlice}(c).
For $1-p \approx 0$  there is a steep increase from the isotropic globule state $\avz \approx 0$ to the crystal phase $\avz \approx 1$.
As $1-p$ increases the sharp change near the transition disappears and the limiting value for large $\eta$ also reduces significantly.
In order to focus on the effect of impurity density on the anisotropy in the collapsed phase we look more closely at the low-temperature (high $\eta$) limiting values of $\avz$.
Because the anisotropy \eref{eq:Anisotropy} is a function of the global structure of each walk we should also account for the finite size of the walks by comparing the relevant length-scales, namely the size of the chains and the mean separation of impurities.
As such we define a scaling parameter $\chi = \sqrt{n \, (1-p)}$, which follows from the mean-squared radius quantities that scale as $\avR \sim n^{2\nu}$, where $\nu = 1/2$ for the collapsed phases in two dimensions.
\fref{fig:LimitingOrderParameters} shows the values of $\avz$ at a large value of $\eta$ representing the low-temperature limit, namely $\eta=5$, as a function of $\chi$.
The values of $\chi$ are achieved by taking data from a several values of $n$ for each value of $1-p$.
As seen before, for small $\chi$ the limiting value of $\avz$ is close to unity indicating the crystal phase.
At large values of $\chi$ the limiting value of $\avz$ is reduced, but this appears to be more a function of $1-p$ only rather than $n$ via $\chi$. This is the same behaviour as $\avs$ and $\avm$.
However, there is an intermediate regime \smash{$5 \lesssim \chi \lesssim 8$} where the limiting value of $\avz$ decreases sharply.
This indicates that at some critical amount of inhomogeneity the anisotropic configurations which typify the crystal phase are inhibited.
The only remaining distinction from the globule phase is that $\avz > 0$ for large $\chi$, or at least it tends to zero slowly, reflecting that the smaller sub-clusters retain some anisotropic order, but this is uncorrelated with other sub-clusters on the scale of the entire walk, as per \fref{fig:Configurations}(h).

Taken together, the behaviour of the order parameters indicate that as disorder, measured by $\chi$, increases the crystal phase disappears.
Within the remaining collapsed phase there is still a dependence on $\eta$, i.e.~stiffness.
At one end, $\eta = 1$, the system is most like the globule phase from the homogeneous lattice case where only $\avm$ decreases as $1-p$ increases, representing the formation of clusters at a smaller scale.
As $\eta$ is increased to where the crystal phase was, the overall anisotropy remains small but $\avs$ increases, albeit to a smaller value than the homogeneous lattice case.
This indicates that there is an increase of stiffness within each smaller cluster but with no correlation between the smaller clusters since the impurities are distributed uniformly.
Furthermore, since there is no sharp change in any of the order parameters at some intermediate $\eta_c$ then the transition between these regimes is smooth and not critical. 
Thus, the globule-crystal transition has disappeared and any remnant indications of a transition that appear in \fref{fig:HessianPhase}(b) are therefore argued to be due to finite-size effects, such as conformational rearrangements, rather than due to the finite-size rounding of a thermodynamic transition.

\section{Conclusion}
\label{sec:Conclusion}

We have simulated semi-stiff ISAWs on the square lattice with variable amount of lattice inhomogeneity represented as impurities, or sites that are unavailable to the walks.
We have characterised the three expected phases, extended, globule and crystal, and looked at how the latter two collapsed phases are affected by increasing the inhomogeneity.
When a large number of impurities are present the walks prefer a configuration of several smaller clusters as opposed to a global swelling.
This is consistent with other studies of long chain polymers in disordered media \cite{Goldschmidt2003}.
For finite length walks the effect on the crystal phase is strongest, where we observe changes to the order parameters.
In particular, there is a significant decrease in the average anisotropy as well as a softening of the crystal-globule transition.
The latter is seen as a rounding of the singular behaviour in $\avs$ to the point where it is no longer clearly distinguishable as a critical transition.
We conclude that the crystal phase and the critical transition between the crystal phase and globule phase disappear when the mean distance between impurities is comparable to the size of the chains, as judged by the scaling of mean-squared radius $\avR$. 
We identify that the ratio of these length scales in the system $\chi$, or rather $\chi^2 = n \, (1-p)$, serves as the appropriate scaling variable near the onset of disorder in the lattice, i.e.~$1-p > 0$.
Hence in the thermodynamic limit there is no crystal phase for any amount of inhomogeneity.
We also note that the scaling variable depends on the length-scale exponent $\nu = 1/2$ and thus the softening of the transition is consistent with the Harris criterion for the effect of quenched disorder on critical phenomena \cite{Harris1974}.

Beyond the exploration of the phase diagram the natural question is to look at the scaling behaviour of the order parameters, fluctuations of these and size measures in each of the phases and also at the critical transitions between them.
While the machinery of finite-size scaling has proved to be powerful for other SAW problems, in this case the data gathered is not good enough to be used for such analysis.
In particular we are unable to locate critical points with sufficient precision.
The addition of disorder will always present a challenge to computational methods but in this case the stronger effect is due to the challenge of simulating dense configurations with enough fidelity to distinguish between different behaviours in the collapsed phase.

\begin{acknowledgements}
Financial support from the Australian Research Council via its Discovery Projects scheme (DP160103562) is gratefully acknowledged by the authors. 
\end{acknowledgements}




\end{document}